\documentclass[doublecol,linenumbers]{epl2}  
\usepackage{amssymb}
\usepackage{bm}
\usepackage{amsmath} 
\usepackage{graphicx}
\usepackage{epsfig}
\usepackage{dsfont}

\RequirePackage{color}
\definecolor{MyDarkGreen}{rgb}{0.02,0.60,0.06}
\definecolor{burgundy}{rgb}{0.5, 0.0, 0.13}

\newcommand\be{\begin{equation}}
\newcommand\ee{\end{equation}}
\newcommand\beq{\begin{eqnarray}}
\newcommand\eeq{\end{eqnarray}}

\def\balpha{{\mbox{\boldmath $\alpha$}}} 
\def\bsigma{{\mbox{\boldmath $\sigma$}}} 
\def\bnabla{{\mbox{\boldmath $\nabla$}}} 
\def\vec#1{{\bf #1}}

\begin{document}
\date{\today}

\title{{Using torsion to manipulate spin currents}}

\author{S\'ebastien Fumeron\inst{1}, Bertrand Berche\inst{1,2}, Ernesto Medina\inst{1,2,3}, Fernando A. N. Santos\inst{4} \and Fernando Moraes\inst{1,5,6}}
\shortauthor{S. Fumeron \etal}
\institute{
  \inst{1}{ Statistical Physics Group, IJL, UMR Universit\'e de Lorraine - CNRS 7198,
    54506 Vand\oe uvre les Nancy, France}\\
   \inst2{Centro de F\'isica, Instituto Venezolano de Investigaciones Cient\'ificas, 21827, Caracas, 1020 A, Venezuela}\\
   \inst3{Yachay Tech, School of Physical Sciences \& Nanotechnology, 100119-Urcuqu\'i, Ecuador} \\
   \inst4{Departamento de Matem\'atica  Universidade Federal 
de Pernambuco, 50670-901, Recife, PE, Brazil } \\                   
  \inst{5}{Departamento de F\'{\i}sica, CCEN, Universidade Federal da Para\'{\i}ba,
 58051-900, Jo\~ao Pessoa, PB, Brazil }\\
  \inst{6}{Departamento de F\'{\i}sica, Universidade
    Federal Rural de Pernambuco, 
    52171-900 Recife, PE, Brazil}}

\abstract{We address the problem of quantum particles moving {on} a manifold {characterised} by the presence of torsion along a preferential axis. {In fact, such a torsion may be taylored by the presence of a single screw dislocation, whose Burgers vector measures the torsion amplitude.} The problem, {first treated in the relativistic limit describing fermions that couple minimally to torsion, is then {analysed} in the {Pauli limit}}. We show that torsion induces a geometric potential and also that it couples generically to the phase of the wave function, giving rise to the possibility of using torsion to manipulate spin currents in the case of spinor wave functions. These results emerge as an alternative strategy for using screw dislocations in the design of spintronic-based devices.}

\date{\today}
\maketitle

\section{Introduction}
{The first systematic studies on topological defects in solids began with Volterra \cite{volterra1907equilibre} in the beginning of the last century,
but it was only relatively recently that it was shown that the deformation field {couples these defects to} electron spin\cite{Schlenker} and cause 
phenomena such as magnetoelasticity\cite{eringen2012electrodynamics}.}
Spintronics, on the other hand, is quite a recent subject\cite{ReviewSarma} and surprisingly, few studies relate spin currents to defects. Such studies, in general, blame defects for compromising spin transport. Counter to this argument, in this letter, we  propose that topological defects, like dislocations, might be used as channels for spin transport due to the unique geometry associated that involves torsion, but not curvature. {In spite of the fact that dislocations in crystal are mobile and thermally activated, we will restrict ourselves to the case of an isolated fixed dislocation. For potential practical applications, this may be controlled by pinning points created by point
defects, substitutional impurities or grain boundaries, depending on the material, in order to reduce the mobility of the defect (which should otherwise overcome pinning energy). The pinning defects must not couple to spin so that no degradation of the
spin currents occurs. }

{The coupling of matter to geometry through curvature, was brilliantly demonstrated by Einstein's general theory of relativity, and was then extended, in Einstein-Cartan's approach, to  couple the spin density tensor and torsion \cite{RevModPhys.48.393} (see also the review on torsion by Hammond~\cite{Hammond02} 
for a very nice and personal account). While the order of magnitude of this latter coupling in astrophysical applications is possibly too small to be of practical interest, the situation might be drastically different in the condensed matter physics arena. This letter is intended to explore the possible consequences that emerge for spintronics applications.}

\begin{figure}
\includegraphics[width=80mm]{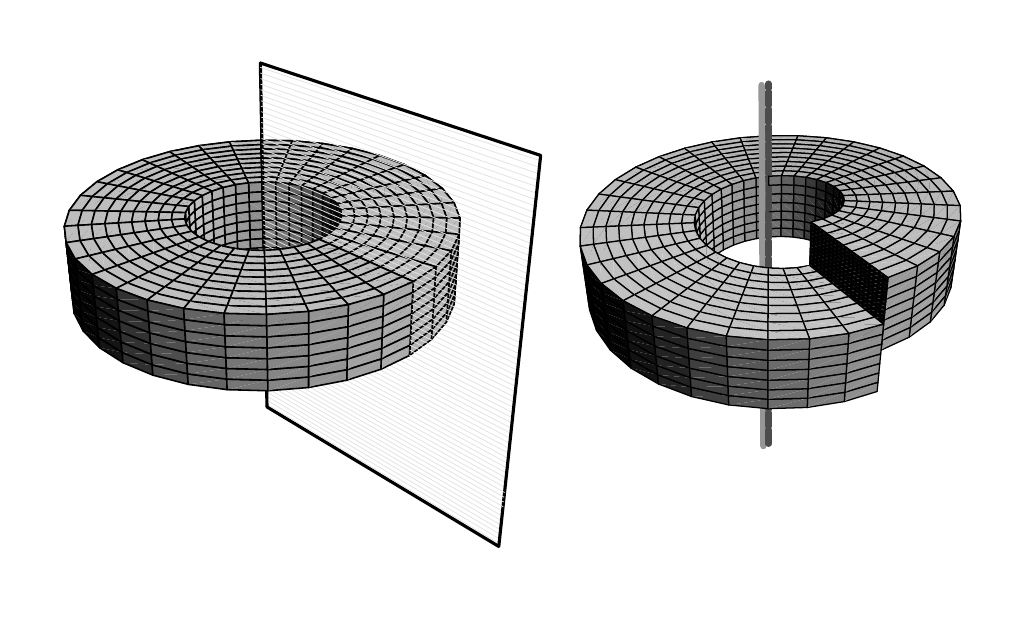}\vspace{-8mm}
\caption{Volterra process for a screw dislocation:  cut, {displace} and glue (adapted from  \cite{puntigam1997volterra}).}
\label{fig.0}
\end{figure}

Dislocations are a result of discrete translational symmetry breaking in otherwise perfectly periodic crystals. Their association with translational broken symmetry can easily be grasped in a topological ``cut and glue" process known as Volterra process\cite{friedel1967dislocations}. Even though crystals have a discrete structure, if a quantum particle has low enough energy, it cannot perceive the lattice detail and therefore an effective continuum theory, incorporating topological properties of the original crystal, can be used\cite{RevModPhys.65.733}. 
The Volterra process for constructing a screw dislocation is sketched in Fig. \ref{fig.0}. It follows that the Euclidean line element $ds^2 =dr^2 + r^2 d\phi^2 + dz^2$ is changed to
\begin{equation}
ds^2 =dr^2 + r^2 d\phi^2 + (dz+ \kappa d\phi)^2, \label{metric}
\end{equation}
where $\kappa = b/2\pi$ and $\vec{b}=(0,0,b)$ is the Burgers vector associated to the defect. Since the metric given by Eq.(\ref{metric}) was obtained by a rigid displacement and subsequent glueing of the surfaces, without allowing the solid to relax, {an infinite stress is associated to the axis, resulting in a torsion singularity}. 
Geometrically, a dislocation in a rigid crystalline structure can be seen as a torsion vortex in the continuum limit\cite{katanaev1}, with the corresponding $\delta$-function singularity at the defect line, as described by Puntigam and Soleng\cite{puntigam1997volterra}. These authors characterized the geometry associated to the ten possible Volterra line defects in {four} space-time dimensions. This included screw and edge dislocations, as well as space-time defects like cosmic strings. Even though an edge dislocation would also be of interest for our discussion, for simplicity, we choose the screw dislocation geometry, whose only non {zero} component of the torsion tensor is\cite{puntigam1997volterra}
{$T_{zxy} = \kappa  \delta(x)\delta(y)$. 
This equation is then} the torsion vortex of a screw dislocation in a rigid crystal associated to {an} infinite stress on the axis. On the other hand, real crystals relax distributing the stress around the defect axis and therefore broadening the singularity into a smooth function. Sharma and Ganti\cite{sharma2005gauge} and Lazar and Anastassiadis\cite{lazar2009gauge} studied the gauge theory of dislocations in the relaxed solid. The resulting torsion for the screw dislocation
is\cite{lazar2009gauge}
\begin{eqnarray}
T_{zxy} = \frac{\kappa }{ r_0^2}K_{0}(r/r_0) , 
\label{torsionK}
\end{eqnarray}
where  $K_0$ is the modified Bessel function of the second kind of order zero and $r_0$ is a characteristic {core} length scale of the defect.

Torsion, in three spacial dimensions, has {nine} independent components and can be written  as the sum of its irreducible parts \cite{McCrea}: tensor, trace and axial . In this case we {only} have axial torsion, called \textit{axitor}, whose unique component is {in the direction of the Burgers vector, and follows} from Eq. (\ref{torsionK}),
\begin{equation}
A^z = \frac{\kappa }{ r_0^2}K_{0}( r/r_0)\label{S3}.
\end{equation}
Having set the background geometry associated to the defect, we will describe in the next section how torsion can affect the motion of free spins and point out the possibility of application to spintronics

\section{Interaction of torsion with a spin 1/2 particle}
\subsection{General problem}
The deformation field of a dislocation generates an effective geometry that {influences} classical trajectories and quantum states of spinning particles moving in the {vicinity} of the dislocation. In fact, torsion can couple to the spin of the particles in the same way an applied external magnetic field. This coupling is obtained by writing the action for a fermion field in the presence of gravity with torsion\cite{Hammond02,shapiro,Zecca2002}. 
{{{While different choices are found in the literature, here we adopt} what we believe is the most natural: The free particle Dirac equation takes the form $(i\gamma^\mu\partial_\mu-m)\psi=0$ with a Minkowskian signature $\eta_{\mu\nu}=(+,-,-,-)$, yielding the algebra $\{\gamma^\mu,\gamma^\nu\}=2\eta^{\mu\nu}$. The coupling of fermions to torsion, in absence of magnetic field, is described by the action \cite{bagrov2}
\be
S=\int d^4x(\bar\psi\gamma^\mu(i\partial_\mu-\eta_1\gamma_5A_\mu-\eta_2T_\mu-m)\psi,
\ee 
with $\bar\psi = \psi^{\dagger}\gamma^0$, and  $A_\mu=(A_0,-\vec A)$, $T_\mu=(T_0,-\vec T)$ are the axial and trace parts of torsion, respectively. The associated Dirac equation in standard form is 
\be i\partial_t\psi=[\balpha\cdot(\vec p-\eta_1\gamma_5\vec A-\eta_2\vec T)+\eta_1\gamma_5A_0+\eta_2T_0+\beta m]\psi,\label{EqDirac}\ee 
where $\balpha$ and $\beta$ are the usual Dirac matrices and here $\gamma_5=\gamma^5=i\gamma^0\gamma^1\gamma^2\gamma^3$. 
This choice of action is natural in the sense that the terms are written in a similar manner as the coupling to an electromagnetic field: in the Dirac equation this choice adds $-e\vec A$ to $\vec p$ and $e\phi$ to $H$ with $e<0$.
We {note here that some choices in the literature present inconsistencies, which may result in non hermiticity or {erroneous} signs}.  The non relativistic limit, {to lowest order} follows as
\beq H&=&\frac{1}{2m}[(-i\bnabla-\eta_2\vec T)+\eta_1 A_0\bsigma]^2 +m\nonumber\\ &&-\frac{\eta_2}{2m}\bsigma\cdot(\bnabla\times\vec T)-\frac{\eta_1^2}{m}A_0^2-\eta_1\bsigma\cdot\vec A+\eta_2T_0,\ 
\label{EqPauli}\eeq where 
$\bsigma$ is the vector of Pauli matrices.
The minimal coupling prescription requires $\eta_1=-1/8$ and $\eta_2=0$.}
As it happens, fermions couple to the axial and trace parts of torsion \cite{OConnor,Buchbinder,bagrov2,shapiro,Audretsch81,Hammond95,Hammond95a,Hammond02,Zecca2002}, 
while $A_0$ is the time-axis component of the torsion axial quadrivector and $\vec{A}$ represents its space part. The 4D torsion  axial vector, which corresponds to the 3D \textit{axitor} of the previous section, is defined as
\begin{equation}
A^{\nu} =\epsilon^{\alpha\beta \mu \nu}T_{\alpha\beta \mu}, \label{S}
\end{equation}
where $T_{\alpha\beta\mu}$ is the 4D torsion tensor. 

In the non-relativistic limit, after relabeling $\eta_1=\eta$ as the {effective} coupling constant between torsion and matter fields, the Dirac operator (\ref{EqDirac}) results in the low-energy Pauli Hamiltonian  
\begin{equation}
H= \frac{1}{2m}\left(\vec{p}+\frac{\eta}{c} A_0 \bsigma \right)^2 - \frac{1}{mc^2}\eta^2 A_0^2 {- \eta c\bsigma \cdot  \vec{A}} . \label{H}
\end{equation}
The time-like part of torsion, $A_0$, couples to the electron spin via a minimal coupling to the kinematic momentum in Eq.~(\ref{H}) in a manner analogous to the way spin-orbit interactions can be incorporated into the kinetic energy through a non-Abelian gauge field in the Pauli equation \cite{JinLiZhangJPA,MedinaLopezBercheEPL,BercheMedinaEJP}. The presence of the second term in Eq.~(\ref{H}) plays the role of a mass term for torsion (it is quadratic in $A_0$) \cite{BercheMedinaLopezEPL}. It adds to the gauge invariant form of the minimally coupled term of the kinetic energy\cite{BercheMedinaLopezEPL}.  Finally, the last term is analogous to a Zeeman interaction which couples the spatial part of torsion to the spin and will be discussed below.
It is worth noting that the coupling constant of spin to torsion through both the spin-orbit-like and the Zeeman-like terms is characterized by a unique parameter $\eta$, {and both terms offer the opportunity {for} spin manipulations}. 

In the case studied here, torsion is purely spatial, which makes $A^0=0$. It follows from Eq. (\ref{H}), that the Hamiltonian reduces {to the simple form}
\begin{equation}
H= \frac{1}{2m}\vec{p}\,^2  {- \frac{2\eta c}{\hbar} \vec{S} \cdot \vec{A}} , \label{Htor}
\end{equation}
where $\vec{S}=\frac{\hbar}{2}\bsigma$ is the spin angular momentum of the electron.  
The material-dependent coupling constant $\eta$ expresses the magneto-elastic interaction; that is, it describes {the strength of elastic torsion}, due to the defect distribution, on the   particle's spin and should be determined experimentally for each material. Note that $\eta$ has the dimensions of {an action}.  In the case of {the} minimally coupled Dirac spinor in an external torsion field, as we mentioned before, Shapiro and co-authors have shown that {$\eta=(-1/8)\hbar $} and they also considered the possible generalization to other situations\cite{shapiro2,Ryder199821}. For the remaining of this paper, we will keep in mind that $\eta$ is negative and of the order of {unity in units of $\hbar $}.

\subsection{Classical behaviour}

Let us consider a ``classical" particle with spin moving in the presence of the screw dislocation. Following Bagrov, Buchbinder and Shapiro\cite{bagrov2,shapiro2}, the quasi-classical equations of motion, corresponding to the Hamiltonian given by Eq.(\ref{Htor}), are written as: 
\begin{eqnarray}
m\frac{d\vec{v}}{dt} & = & {+ \frac{2\eta c}{\hbar}} \bnabla\left(\vec{S}\cdot \vec{A} \right), \label{eqmov1} \\
\frac{d\vec{S}}{dt} &  = & {- \frac{2\eta c}{\hbar} }\left( \vec{A}\times \vec{S}\right). \label{eqmov2}
\end{eqnarray} 
Note the striking similarity between Eqs. (\ref{eqmov1}) and (\ref{eqmov2}) and the corresponding equations of motion for a magnetic dipole in a magnetic field, which describe both the magnetic force on the dipole due to a non-homogeneous field and the torque which leads to Larmor precession, respectively. Incidentally, the magnetic equivalent of Eq. (\ref{eqmov1}) has been used to {levitate} small diamagnetic objects (including frogs!) in a strong, non-homogeneous, magnetic field \cite{frog1,frog2}. 

Since {$\eta <0$}, according to Eq. (\ref{eqmov1}), the {most favorable spin orientation is {anti}parallel to the Burger's vector and the force on the spin will be attractive when the spin is parallel to {it}, if the gradient of torsion is negative, and repulsive otherwise}. This gives the screw dislocation the ability to attract spins with a specific polarisation. If the captured  carriers are able to move along the defect, this gives rise to a spin-polarized current, {an important resource for spintronics}. In what follows, we analyse the classical motion of the carriers attracted by the defect.

\begin{figure}
\includegraphics[width=95mm]{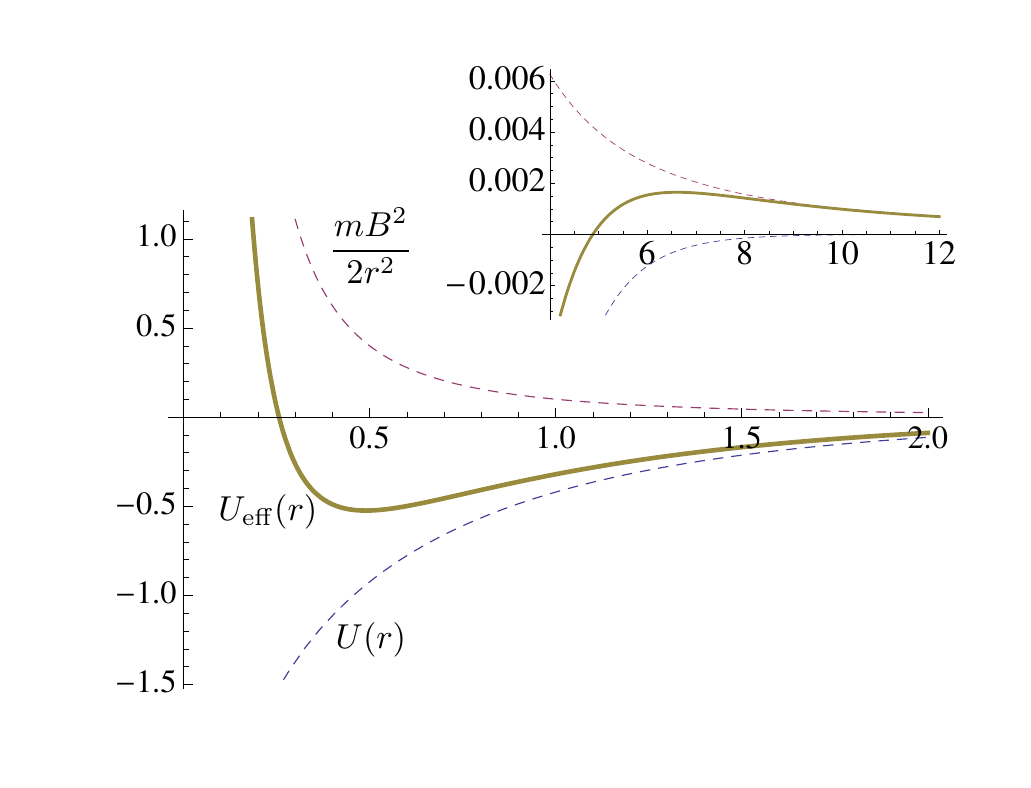}
\caption{Plot of $U(r)=-\alpha K_0 ( r/r_0)$. The potential is in units of $\alpha$ and the radial distance in units of $r_0$. The upper dashed line is the centrifugal potential (with $B=0.447$), the lower one is the potential $U$ (with $\alpha =1$) and the thick solid line is the effective potential $U_{\rm eff}(r)$ that can accommodate bound states. The upper inset shows the local maximum at larger distances. }
\label{fig.1}
\end{figure}

Care must be taken in solving Eqs. (\ref{eqmov1}-\ref{eqmov2}) since the background geometry is not Euclidean. We can get insight into the classical motion {by considering the geometry to be that of the approximate metric} (see Eq. (\ref{metric})).  Dividing $ds^2$ by $dt^2$, we get the velocity squared 
$v^2 = \dot{r}^2 + r^2 \dot{\phi}^2 + (\dot{z}+ \kappa \dot{\phi})^2$,
that leads to the Lagrangian 
\begin{equation}
L = \frac{1}{2}m \left[ \dot{r}^2 + r^2 \dot{\phi}^2 + (\dot{z}+\kappa \dot{\phi})^2 \right] - U(r), \label{lag}
\end{equation}
where the attractive potential 
\begin{equation}
U(r) = -{|\eta| c} \frac{\kappa }{ r_0^2}K_0( r/r_0) = -\alpha K_0(r/r_0) \label{U(r)},
\end{equation}
comes from Eqs.  (\ref{S3}) and (\ref{eqmov1}), where $S$ was replaced by its eigenvalue {$-\hbar/2$}, considering the polarisation {antiparallel} to the Burgers vector. A representation of this potential can be seen in Fig.\ref{fig.1}. 
From the Lagrangian (Eq. \ref{lag}), we finally have the following equations of motion:
\begin{eqnarray}
& \ddot{r}  -  r \dot{\phi}^2 + \frac{1}{m} \frac{d U(r)}{d r}= 0 ,\\ \label{eqr}
& \ddot{\phi} +  \left(\frac{2}{r} \right) \dot{r}\dot{\phi}=0 , \\ \label{eqphi}
& \ddot{z}  +  \kappa\ddot{\phi} =0 , \label{eqz}
\end{eqnarray}
that, without the potential term, coincide with the geodesic equations associated to the screw dislocation geometry  \cite{padua}. Following this reference, we see that eqs. (\ref{eqr}-\ref{eqz}) can be integrated to
\begin{eqnarray}
\dot{r}^2 + \frac{B^2}{ r^2} + \frac{2}{m}U(r) = A ,  \label{eqrr}\\
r^2 \dot{\phi} = B , \label{eqphiphi}\\
z + \kappa \phi = Ct + D , \label{eqzz}
\end{eqnarray}
where $A$, $B$, $C$ and $D$ are integration constants, {$\frac{1}{2}m(A + C^2)$} is the energy and $mB$ is the angular momentum). From Eq. (\ref{eqzz}) we see that $C$ is a conserved, mixed momentum, involving  $\dot{z}$ and $\dot{\phi}$, a peculiarity of this geometric background with torsion.

\begin{figure}
\includegraphics[width=90mm]{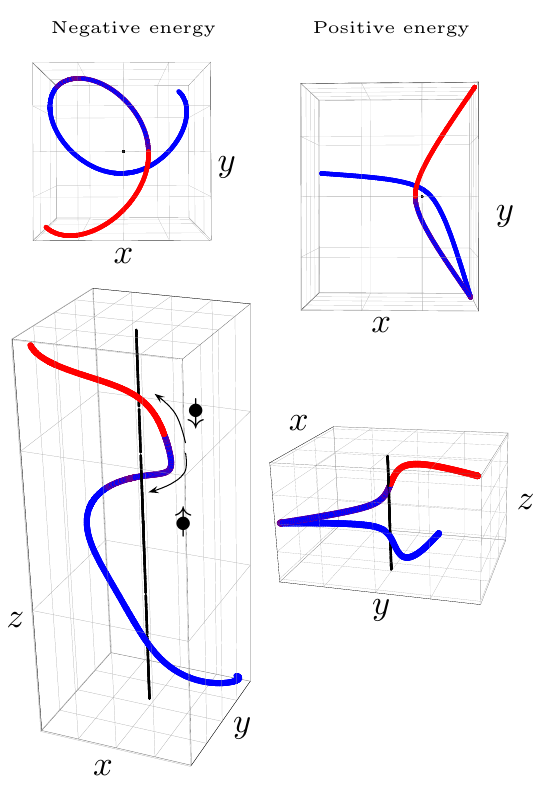}
\caption{Trajectories around the defect line (straight line along $z-$axis) for a bound state of total energy $A=-0.5$ (left) and an asymptotically free state $A=0.5$ (right). For $A=-0.5$ the trajectories are bounded in the perpendicular plane (upper pannel), but the potential does not allow for closed orbits.  The trajectories follow channels along the defect (lower pannel), downwards when the angular momentum is positive ($B=+0.447$, blue part, symbolized by the uparrow) and upwards otherwise ($B=-0.447$, red part). A particle orbiting in the perpendicular plane will automatically scatter along the defect in one direction or the other, depending on the particle's angular momentum, leading to {spin and} orbital momentum-dependent conducting channels. For $A=0.5$, and more generally for free states, the defect can be used to select orbital momentum in the $z-$direction {lower pannel}, which is a peculiarity of spaces with torsion.
}
\label{fig.2}
\end{figure}

The formal solution to Eq. (\ref{eqrr}) is 
\begin{equation}
t = t_0 + \int \frac{dr}{\sqrt{A - B^2/r^2 + (2\alpha/m) K_0( r/r_0) }}, \label{t}
\end{equation}
(where $\alpha$ is defined in Eq.~(\ref{U(r)}))
which, when inverted, gives $r(t)$. With this  {result}, one can then integrate Eq. (\ref{eqphiphi})  for $\phi(t)$ and, finally obtain $z(t)$ from Eq. (\ref{eqzz}). 
The only possibility of attraction is if the effective potential $U_{\rm eff}(r)=\frac{mB^2}{2 r^2} + U(r) <0$ (note that the effective potential can display, for specific values of the parameters, a local maximum at large values of $r$, which slightly tempers this statement). But from the limiting forms of $K_0(x)$  \cite{nist}, that is,  $-\ln x$ when $x \rightarrow 0$ and $\sqrt{\pi/2x}e^{-x}$ as $x \rightarrow \infty$, it is clear that there can be attraction in the plane perpendicular to the defect line [Eq. (\ref{eqzz}) implies that the linear momentum along $z$ is conserved. Hence, there is no bound state in the direction along the defect].
Examples of ``attractive''  (bound state) and asymptotically free trajectories are depicted in Fig. \ref{fig.2}.
The classical picture may then be summarized as follows: i) the defect line selectively captures in the positive $z-$direction particles with spin polarized antiparallel to the Burgers vector ({as found in reference\cite{MedinaJCP}} {a helicity}); ii) the particles will not be captured into orbits around the defect but are rather directed to the defect core and, {depending on their angular momentum as explained in the caption of Fig. \ref{fig.2},} propagated along the defect which acts as a conducting channel; iii) asymptotically free particles are just scattered out of their initial plane of injection towards the defect, this latter source of torsion acting more as a spin splitter in this case (in the sense of a Stern-Gerlach beam splitter\cite{VarelaBornScatt}).

\subsection{Quantum behaviour}

From Eqs. (\ref{Htor}) and (\ref{U(r)}), keeping in mind that the Laplacian operator in non-Euclidean space must be replaced by the Laplace-Beltrami operator $\nabla^2_{LB}= \frac{1}{\sqrt{g}}\partial_i (g^{ij} \sqrt{g}\partial_j)$. We write the Schr\"odinger equation as
\begin{eqnarray}
H\psi = & - & 
\frac{\hbar^2}{2m}\left[\partial^2_z + \frac{1}{r} \partial_r (r \partial_r) + \frac{1}{r^2}\left( \partial_{\phi} - \kappa \partial_z \right)^2 \right]\psi \nonumber \\
& - & 
\alpha K_0( r/r_0) \psi = E \psi , \label{Schr}
\end{eqnarray}
where the background geometry assumed was that given by the metric (Eq. \ref{metric}).  {For the wavefunction}, we use the \textit{ansatz} 
\begin{equation}
\psi (r,\phi,z) = e^{ikz} e^{i\ell \phi} R(r),
\end{equation}
where, in principle, $k \in \mathbb{R}$ and $\ell \in \mathbb{Z}$. 
A comment on the boundary conditions is suitable here: We consider a metric with a source of torsion on the axis, and choosing integer $\ell$'s, we fix periodic boundary conditions for the angular dependence. On the other hand, one knows that a change of coordinate may be performed in order to render the metric just the one of the ordinary  cylindrical coordinates  \cite{DeLorenci}. Nevertheless, going from one geometry to the other is equivalent to performing a gauge transformation and the phase of the wave function is thus modified. The situation is very similar to the discussion of the more usual context of a charged particle in the presence of a magnetic field  {discussed} in  \cite{BercheMalterreMedina}. Let us summarize this argument in the case of the simpler problem of a quantum particle constrained to move at a fixed radial distance $r$, like the one analysed in ref. \cite{DeLorenci}. The  Schr\"odinger equation (Eq. \ref{Schr}) reduces to 
\begin{equation}
 -  \frac{\hbar^2}{2m}\left[\partial^2_z +  \frac{1}{r^2}\left( \partial_{\phi} - \kappa \partial_z \right)^2 \right]\psi  = E \psi . \label{Schrprime}
\end{equation}
Looking for solutions of the form $\psi(\phi,z)=A e^{ikz} e^{i\ell \phi}$, one gets the eigen-energies $-(2m/\hbar^2)E=k^2+(\ell-\kappa k )^2/r^2$ and the periodic boundary conditions $\psi(\phi+2\pi,z)=\psi(\phi,z)$ require $\ell\in\mathbb{Z}$.
De Lorenci and Moreira  \cite{DeLorenci} propose a change of coordinates to write the problem on a flat cylinder instead, $(r,\theta,z)$. Now, 
 $-  \frac{\hbar^2}{2m}\left[\partial^2_z +  \frac{1}{r^2} \partial_{\theta}^2 \right]\psi'  = E \psi' $, 
with $\psi'(\theta,z)$ a gauge transformed wave function describing the same physics, i.e. in particular with the same energies. Writing $\psi'(\theta,z)=A e^{ikz} e^{i\ell' \theta}$, one obtains $-(2m/\hbar^2)E=k^2+{\ell'}^2/r^2$ with $\ell'=\ell-\kappa k\in\mathbb{R}$
and the boundary conditions in the new gauge, in terms of the original angular coordinate become $\psi'(\phi+2\pi,z)=e^{-i2\pi\kappa k}\psi'(\phi,z)$. Note that $\psi'$ is obtained from $\psi$ via a singular gauge transformation  \cite{BercheMalterreMedina}. Contrary to the conclusion of Ref. \cite{DeLorenci}, we see that there are two approaches of the same problem in the presence of torsion: either torsion is explicitly present in the Hamiltonian and the wave functions used are periodic, or torsion is absent from the differential equation, but still appears in the boundary conditions. Hence, torsion's effects on the wave function are made explicit. Both situations describe the same physics, but the wave function is multivalued in the second case. Among other consequences, this necessitates a proper redefinition of the canonical momentum  \cite{Riess}.  The two approaches have been discussed in analogy to respectively Einstein's and Weyl's theories of gravitation in ref. \cite{BercheMalterreMedina}.

Let us now return to  Eq. (\ref{Schr}) with periodic boundary conditions as initially  {required}. The radial equation reduces to
\begin{eqnarray}
\left[ \frac{1}{r}\frac{d}{dr}\left( r\frac{d}{dr}\right) - \frac{(\ell -\kappa k)^2}{r^2} + \frac{2m\alpha}{\hbar^2} K_0( r/r_0)\right] R  \nonumber \\
+ \left(\frac{2m}{\hbar^2}E - k^2 \right)R =0 . \label{RSchr}
\end{eqnarray}

An argument for the possibility of bound states and then an estimate of the ground state  (GS) energy is given via a simple analysis.  {For the} short distance limit $K_0(r/r_0)\sim -\ln(r/2r_0)-\gamma$ with $\gamma$ the Euler constant, and the total energy  takes the form $E=p^2/2m+\alpha\ln(r/2r_0)+\gamma$. It has a minimum at $r=\hbar/\sqrt{m\alpha}$ where the GS energy is then of order $E_0=\alpha(1/2+\ln(\hbar/2r_0\sqrt{m\alpha})+\gamma)$. Bound states can exist only if the minimum distance scale  {is such that} $\hbar/\sqrt{m\alpha}<0.681 r_0$.

As noticed in the previous subsection, classically, the attraction by the dislocation depends on the angular momentum and the latter is intimately linked to the linear momentum along the defect core. We thus have in the quantum context the interesting result that the linear momentum along the $z$-axis must be quantized, $\hbar k=\ell\hbar/\kappa$, for the same reason.
 This also appears when we have an electromagnetic wave propagating along a screw dislocation  \cite{fumeron2015generation}, or similarly, in the helicity dependence of the propagation of heat flow   \cite{Fumeron201364}.  We also  {emphasize} that due to this relation, the currents $j^z$ (e.g. charge or spin) along the $z-$direction will be proportional to the corresponding angular currents $j^\phi$   \cite{SinhaEPJB,MedinaJCP}.
 Taking the quantization of $k$ into account and making the transformation $R(r)=u(r)r^{-1/2}$,  the resulting equation 
\begin{eqnarray}
\frac{d^2 u(r)}{dr^2}  +\frac{2m}{\hbar^2} \left[ E- \frac{\ell^2 \hbar^2}{2m\kappa^2}  + \frac{\hbar^2}{8mr^2}
+ \alpha K_0\left( \frac{r}{r_0}\right) \vphantom{\frac{2m}{\hbar^2}}\right]  u(r)=0,
\end{eqnarray}
also appears  in the study of the two-dimensional hydrogen atom with Maxwell-Chern-Simons interaction, addressed by Caruso \textit{et al.}  \cite{Caruso2013}. In this reference, the authors solve the equation numerically and find results for the energy and average radius of the ground state for three different characteristic length  scales, analogous to our $r_0$.  In Table \ref{table} we reproduce their results. Even though their coupling constant may be quite different from ours, a qualitative comparison is worthwhile. For convenience, we write their length  scale in terms of the Bohr radius   $a_0 = 5.3 \times 10^{-11} $m.
\begin{table}
\centering
\begin{tabular}{| c | c | c| c |}
\hline
$r_0/a_0$ & $\langle r\rangle/a_0$ & $\langle r\rangle/r_0$ & $E_0$ (eV) \\
\hline
36 & 10.65 & 0.30 &- 0.09 \\
\hline
360 & 25.59 & 0.071 & - 0.022 \\
\hline
3600 & 92.35 & 0.026 & -0.0035 \\
\hline
\end{tabular}
\caption{Characteristic length scales $r_0$ and the corresponding mean radii and binding energies for the ground state, taken from ref.  \cite{Caruso2013}. }
\label{table}
\end{table}
If we assume  $r_0$ to be an estimate of the screw dislocation core radius, we see that  the particle, in its bound state, is always inside the defect core. But, the smaller the core size, the closer  {the particle is to the core boundary}, as seen in the $\langle r\rangle/r_0$ column of Table 1. In fact, assuming $r_0 \sim b \sim$ lattice constant  \cite{hirth1982theory}, the first line of the table seems to be the appropriate scale. For silicon, for example, the lattice constant is $5.4 \times 10^{-10} m \sim 10 a_0$. The binding energy is of the order of that of an exciton, which is quite reasonable, considering that there is no electromagnetic interaction between the particle and the dislocation. Again, we remark that this analysis is limited by the differences in the coupling constants of the two problems.

\subsection{Summary and conclusions}
We have found that a defect with torsion selectively captures into radial bound states those spins polarized antiparallel to the Burgers vector with the $z$ component of the linear momentum given by $\hbar k=\ell h/b$. Note that this rule imposes the condition that, if $b$ and $k$ have the same sign (electrons with a definite helicity), \textit{i.e.}, $\vec{b}=b\hat{\vec z}$ and $\vec{k}=k\hat{\vec z}$, then $\ell >0$ and if $b$ and $k$ have opposing signs ($\vec{k}=-k\hat{\vec z}$), then $\ell <0$. Since the captured spins are polarized antiparallel to  $\vec{b}$,  we have, in principle, a two-way channel with opposing helicity, for spin transport. With the eventual inclusion of the spin-orbit interaction, the channel corresponding to  $\ell < 0$ will be favoured and therefore the dominant spin-polarized current will be in this direction ($-\hat{\vec z}$).  

If the spin carriers move along the defect line, we have a spin-polarized current, but how do we achieve this? Electrical conduction along dislocation lines in semiconductors is quite an old subject. Probably its first experimental observation was in 1953 \cite{PhysRev.93.666} and it is still nowadays the subject of intense research \cite{7422696}. One possible mechanism for dislocation-conduction is the Poole-Frenkel effect  \cite{sze2006physics}, but, whatever the mechanism responsible for the electrical conduction,  the spins of the charge carriers will be polarized, as seen above. The direction of the spin current may also be modified through the deformation of the polarisation vector, as analyzed in Ref.  \cite{Wang2015327}.

In the biological context, an equivalent situation occurs for spin propagation in the presence of spin-orbit interaction via hopping between $p_z$ orbitals of chiral molecules with a carbon skeleton  \cite{MedinaJCP,MedinaPRB}. There, the chirality of the molecule and the direction of propagation along $z$ select the sign of $\ell$.

Since the early 1950's it is known that dislocations in semiconductors may conduct electricity  \cite{PhysRev.93.666}. The advent of spintronics poses the obvious question whether or not dislocations can support spin currents as well. In this letter, we propose a mechanism by which this happens due to the influence of the defect's torsion on the spin carriers, which selects their spin polarisation, resulting in a spin-polarized current along the defect line. Furthermore, the fact that torsion couples to the electron spin suggests the possibility of defect engineering for applications  in spintronics. In the case presented here, a  force associated to the gradient of torsion may be used to bind the spins to the  defect, creating thus a channel for a spin current along the defect axis. The use of torsion gradients to manipulate spin currents is a great advantage over magnetic fields since, while the magnetic field acts on both charge and spin, torsion couples only to spin.   Since the spins polarized in the opposite direction of the Burgers vector will be attracted to the torsion gradient of the defect, one may use different defects carrying torsion (screw dislocation, screw dislocation dipole, edge dislocation, etc.) to get  channels with different characteristics. A screw dislocation dipole, for instance, provides a double channel, for opposite spin polarisations.   {By} the same reasoning, dislocation loops will provide spintronic circuits for polarized spins. 
{The helicity preserving transport requires scattering that is spin active in the same way as edge states
in graphene in the presence of SO interaction. This fact protects spin currents against scattering as long as the potential
fluctuations do not couple to spin.}

In conclusion, torsion due to dislocations seems to provide a pathway for spin circuitry design and spin current control in the absence of a real external magnetic field.

\acknowledgments
F.M. and E.M. are thankful for the financial support and warm hospitality  of the Statistical Physics Group at Universit\'e de Lorraine. This work has been partially supported (F.M.) by CNPq, CAPES and FACEPE (Brazilian agencies). E.M. was partially supported by Yachay tech School of Physics and Nanotechnology and PICS CNRS-FONACIT (Venezuela).

\bibliographystyle{eplbib}

\bibliography{RefsArXiv}

\begin{thebibliography}{10}
\expandafter\ifx\csname url\endcsname\relax\def\url#1{\texttt{#1}}\fi

\bibitem{volterra1907equilibre}
\Name{Volterra V.} \Book{Sur l'{\'e}quilibre des corps {\'e}lastiques
  multiplement connexes} in proc. of \Book{Annales scientifiques de l'Ecole
  Normale superieure} Vol.~24 (Soci{\'e}t{\'e} math{\'e}matique de France) 1907
  pp. 401--517.

\bibitem{Schlenker}
\Name{Kl\'eman M. \and Schlenker M.} \REVIEW{J. Appl. Phys.}{43}{1972}{3184}.

\bibitem{eringen2012electrodynamics}
\Name{Eringen A.~C. \and Maugin G.~A.} \Book{Electrodynamics of continua I:
  foundations and solid media} (Springer Science \& Business Media) 2012.

\bibitem{ReviewSarma}
\Name{Zutic I., Fabian J. \and Sarma S.~D.} \REVIEW{Rev. Mod.
  Phys.}{76}{2004}{323}.

\bibitem{RevModPhys.48.393}
\Name{Hehl F.~W., von~der Heyde P., Kerlick G.~D. \and Nester J.~M.}
  \REVIEW{Rev. Mod. Phys.}{48}{1976}{393}.

\bibitem{Hammond02}
\Name{{R. T. Hammond}} \REVIEW{Rep. Prog. Phys.}{65}{2002}{599}.

\bibitem{friedel1967dislocations}
\Name{Friedel J.} \Book{Dislocations} Addison-Wesley series in metallurgy and
  materials (Pergamon Press) 1967.

\bibitem{RevModPhys.65.733}
\Name{Fr\"ohlich J. \and Studer U.~M.} \REVIEW{Rev. Mod. Phys.}{65}{1993}{733}.

\bibitem{puntigam1997volterra}
\Name{Puntigam R.~A. \and Soleng H.~H.} \REVIEW{Class. Quantum
  Grav.}{14}{1997}{1129}.

\bibitem{katanaev1}
\Name{Katanaev M.~O. \and Volovich I.~V.} \REVIEW{Ann. Phys. - New
  York}{216}{1992}{1}.

\bibitem{sharma2005gauge}
\Name{Sharma P. \and Ganti S.} \REVIEW{Proc. R. Soc. London, Ser.
  A}{461}{2005}{1081}.

\bibitem{lazar2009gauge}
\Name{Lazar M. \and Anastassiadis C.} \REVIEW{Phil. Mag.}{89}{2009}{199}.

\bibitem{McCrea}
\Name{McCrea J.~D.} \REVIEW{Class. Quantum Grav.}{9}{1992}{553}.

\bibitem{shapiro}
\Name{Shapiro I.~L.} \REVIEW{Phys. Rep.}{357}{2002}{113}.

\bibitem{Zecca2002}
\Name{Zecca A.} \REVIEW{Int. J. Theor. Phys.}{41}{2002}{421}.

\bibitem{bagrov2}
\Name{Bagrov V.~G., Buchbinder I.~L. \and Shapiro I.~L.} \REVIEW{arXiv preprint
  hep-th/9406122}{}{1994}{}.

\bibitem{OConnor}
\Name{{M. P. O'Connor and P. K. Smrz}} \REVIEW{Aust. J. Phys.}{31}{1978}{195}.

\bibitem{Buchbinder}
\Name{{I. L. Buchbinder, S. Odintsov, L. Shapiro}} \Book{Effective Action in
  Quantum Gravity} (CRC Press) 1992.

\bibitem{Audretsch81}
\Name{{J. Audretsch}} \REVIEW{Phys. Rev. D}{24}{1981}{1470}.

\bibitem{Hammond95}
\Name{{R. Hammond}} \REVIEW{Class. Quantum Grav.}{12}{1995}{279}.

\bibitem{Hammond95a}
\Name{{R.T. Hammond}} \REVIEW{Phys. Rev. D}{52}{1995}{6918}.

\bibitem{JinLiZhangJPA}
\Name{Jin P.-Q., Li Y.-Q. \and Zhang F.-C.} \REVIEW{J. Phys. A: Math.
  Gen.}{39}{2006}{7115}.

\bibitem{MedinaLopezBercheEPL}
\Name{Medina E., L\'opez A. \and Berche B.} \REVIEW{EPL}{83}{2008}{47005}.

\bibitem{BercheMedinaEJP}
\Name{Berche B. \and Medina E.} \REVIEW{Eur. J. Phys.}{34}{2013}{161}.

\bibitem{BercheMedinaLopezEPL}
\Name{Berche B., Medina E. \and L\'opez A.} \REVIEW{EPL}{97}{2012}{67007}.

\bibitem{shapiro2}
\Name{Shapiro I.~L.} \REVIEW{arXiv preprint hep-th/9811072}{}{1998}{}.

\bibitem{Ryder199821}
\Name{Ryder L.~H. \and Shapiro I.~L.} \REVIEW{Phys. Lett. A}{247}{1998}{21 }.

\bibitem{frog1}
\Name{Berry M.~V. \and Geim A.~K.} \REVIEW{Eur. J. Phys.}{18}{1997}{307}.

\bibitem{frog2}
\Name{Simon M.~D. \and Geim A.~K.} \REVIEW{J. Appl. Phys.}{87}{2000}{6200}.

\bibitem{padua}
\Name{de~Padua A., Parisio-Filho F. \and Moraes F.} \REVIEW{Phys. Lett.
  A}{238}{1998}{153 }.

\bibitem{nist}
\Name{Olver F. W.~J.} \Book{NIST handbook of mathematical functions} (Cambridge
  University Press) 2010.

\bibitem{MedinaJCP}
\Name{Medina E., Gonz\'alez-Arraga L.~A., Finkelstein-Shapiro D., Berche B.
  \and Mujica V.} \REVIEW{J. Chem. Phys.}{142}{2015}{194308}.

\bibitem{VarelaBornScatt}
\Name{Varela S., Medina E., L\'opez F. \and Mujica V.} \REVIEW{Journal of
  Physics: Condensed Matter}{26}{2014}{015008}.

\bibitem{DeLorenci}
\Name{De~Lorenci V.~A. \and Moreira~Jr E.~S.} \REVIEW{Phys. Lett.
  A}{376}{2012}{2281}.

\bibitem{BercheMalterreMedina}
\Name{Berche B., Malterre D. \and Medina E.} \REVIEW{Am. J.
  Phys.}{84}{2016}{616}.

\bibitem{Riess}
\Name{Riess J.} \REVIEW{Helv. Phys. Acta}{45}{1972}{1066}.

\bibitem{fumeron2015generation}
\Name{Fumeron S., Pereira E. \and Moraes F.} \REVIEW{Physica B: Condens.
  Matter}{476}{2015}{19}.

\bibitem{Fumeron201364}
\Name{Fumeron S., Pereira E. \and Moraes F.} \REVIEW{Int. J. Therm.
  Sci.}{67}{2013}{64 }.

\bibitem{SinhaEPJB}
\Name{{D. Sinha}} \REVIEW{Eur. Phys. J. B}{88}{2015}{83}.

\bibitem{Caruso2013}
\Name{Caruso F., Helay{\"e}l-Neto J.~A., Martins J. \and Oguri V.} \REVIEW{Eur.
  Phys. J. B}{86}{2013}{1}.

\bibitem{hirth1982theory}
\Name{Hirth J.~P. \and Lothe J.} \Book{Theory of dislocations} (John Wiley \&
  Sons) 1982.

\bibitem{PhysRev.93.666}
\Name{Pearson G.~L., Read W.~T. \and Morin F.~J.} \REVIEW{Phys.
  Rev.}{93}{1954}{666}.

\bibitem{7422696}
\Name{Couso C., Iglesias V., Porti M., Claramunt S., Nafria M., Domingo N.,
  Cordes A. \and Bersuker G.} \REVIEW{IEEE Electron Device
  Lett.}{37}{2016}{640}.

\bibitem{sze2006physics}
\Name{Sze S.~M. \and Ng K.~K.} \Book{Physics of semiconductor devices} (John
  Wiley \& sons) 2006.

\bibitem{Wang2015327}
\Name{Wang J., Ma K., Li K. \and Fan H.} \REVIEW{Ann. Phys. - New
  York}{362}{2015}{327 }.

\bibitem{MedinaPRB}
\Name{Varela S., Mujica V. \and Medina E.} \REVIEW{Phys. Rev.
  B}{93}{2016}{155436}.

\end{thebibliography}

\end{document}